\documentstyle[prl,aps]{revtex}

\newcommand{\beq}{\begin{equation}}
\newcommand{\eeq}{\end{equation}}
\newcommand{\bea}{\begin{eqnarray}}
\newcommand{\eea}{\end{eqnarray}}
\newcommand{\non}{\nonumber}
\newcommand{\rf}[1]{(\ref{#1})}

\begin{document}
\draft
\title{Quantum Probabilistic Subroutines and Problems in Number Theory}
\author{A. Carlini and A. Hosoya}
\address{Department of Physics, Tokyo Institute of Technology,
Oh-Okayama, Meguro-ku, Tokyo 152, Japan}
    \twocolumn[\hsize\textwidth\columnwidth\hsize\csname
    @twocolumnfalse\endcsname
\maketitle
%


\begin{abstract}

We present a quantum version of the classical probabilistic
algorithms $\grave{a}$ la Rabin.
The quantum algorithm is based on the essential use of Grover's
operator for the quantum search of a database and of Shor's Fourier
transform for extracting the periodicity of a function, and
their combined use in the counting algorithm originally introduced
by Brassard et al.
One of the main features of our quantum probabilistic algorithm is its
full unitarity and reversibility, which would make its use possible
as part of larger and more complicated networks in quantum computers. 
As an example of this we describe polynomial time algorithms for studying 
some important problems in number theory, such as the test of the
primality of an integer, the so called 'prime number theorem' and
Hardy and Littlewood's conjecture about the asymptotic number of
representations of an even integer as a sum of two primes.

\end{abstract}


\pacs{PACS numbers: 03.67.Lx, 89.70.+c, 02.10.Lh}
    \vskip 3ex ]

\narrowtext


\section{Introduction}

Quantum computers allow a superposition of $|0>$ and $|1>$ qubits with
coefficients being complex numbers $\alpha $ and $\beta$,
\begin{equation}
|\psi>= \alpha |0>+\beta|1>. 
\label{SUP}
\end{equation}
It is this superposition which provides us with an enormous number of parallel
computations by generating a superposed state of a large number of terms, 
for example starting with the flat superposition $(|0>+|1>)^N$. \hfil\break
Quantum computers can do  unitary transformations and also make quantum
mechanical
observations which induce an instantaneous state reduction to $|0>$ or $|1>$
with the  probability $|\alpha|^2$ or $|\beta|^2$, respectively 
[1-3].

At present there are two main kinds of interesting quantum algorithms 
which can 
beat their classical counterparts, i.e. Shor's algorithm for factoring 
integers \cite{shor} and Grover's algorithm for the unstructured database 
search \cite{grover} which achieve, respectively, an exponential and square 
root speed up compared to their classical analogues.
One of the most interesting algorithms where these two basic
unitary blocks are exploited in conjunction is the counting algorithm
introduced by Brassard et al. \cite{brassard} (see section 3 for
a detailed description), which can count the cardinality $t$ of a 
set of states with a given property present in a flat superposition
of $N$ states in a time which is polynomial in the ratio $N/t$, and
with an accuracy which can be made exponentially close to one.

In this work we shall show how an extended use of this algorithm 
can be exploited to construct unitary and fully reversible operators
which are able to emulate at the quantum level a class of classical
probabilistic algorithms.
Classical probabilistic algorithms are characterized by the use of random
numbers during the computation, and the fact that they give the correct
answer with a certain probability of success, which can be usually made 
exponentially close to one by repetition (see, e.g., ref. \cite{rabin}).
In this paper we show explicitly 
a series of quantum, fully reversible and unitary
algorithms which can be seen as the quantum analogue of the
aforementioned classical randomized algorithms, in the sense
that they naturally select the 'correct' states with an arbitrarily 
large probability amplitude in the end of the computation, and that
the final measuring process is only an option which may not
be used, e.g., in the case when the 'answer' provided by such
quantum algorithms is needed as a partial (subroutine) result for further 
computations in a larger and more complex quantum network.
The main ingredients for the construction of our quantum algorithms
consist in the repeated use of the quantum counting transform
of ref. \cite{brassard}, the exploitation of the resulting interference 
and entanglement among quantum states and, to some extent, in assigning to 
some extra ancilla qubits the role analogue of the classical random 
repetitions.
Previous work \cite{bennett} (see also ref. \cite{aharonov} for other
types of subroutines in the context of generalized quantum computation)
also dealt with the problem of building unitary and reversible subroutines
for the use in larger quantum computational networks.
However, as quantum interference is not exploited, these methods in general 
require a larger memory space compared to our algorithms (see
also the footnote at the end of section III).

The paper is structured as follows.
In section 2 we summarize the main properties of one of the prototypes
of the classical randomized algorithms, i.e. Rabin's test for primality
of an integer.
In section 3 we describe the main block of our quantum algorithm and,
as a warm up exercise propaedeutical to section 4, we study again
the case of the test of primality for a given integer, comparing
our results with the classical ones.
In section 4 we extend our quantum methods to the problem of
checking the so called 'prime number theorem' concerning the 
distribution of primes smaller than a given integer.
We conclude in section 5 with some discussion and future perspectives.
Finally, in the appendix we suggest how one might test Hardy and
Littlewood's formula concerning the asymptotic behaviour of
the number of representations of a given even integer as the sum
of two primes, and also comment about the possible proof of a famous
Goldbach's conjecture in number theory.  

\section{Classical Randomized Primality Test}

One of the prototype examples of classical probabilistic algorithms
is that of Rabin \cite{rabin} for testing the primality of a given number
$k$.\footnote[1]{For 
a review of other probabilistic and deterministic classical
tests of primality see, e.g., ref. \cite{ribenboim}.}
The algorithm is probabilistic as it uses random integers during the 
computation, it is always correct when certifies a number to be
composite, while it asserts primality with an arbitrarily small 
probability of error.
The algorithm tests the following condition $W_k(a)$, for $1\leq a< k$
(and factoring out of $k-1$ the highest power of 2 which divides it, i.e. 
writing $k-1\equiv 2^h l$, with $h$ integer and $l$ odd):

\bea
(i) &~& a^{k-1} \bmod k \neq 1; 
\non \\
(ii) &~& \exists ~i\in [1, h] ~/~ \gcd(a^{(k-1)/2^i}, k)\neq 1.
\label{1}
\eea

If at least one of conditions (i) or (ii) is satisfied,
then $W_k(a)=0$ and $a$ is said to be a witness to the compositeness of $k$.
On the other hand, if neither (i) nor (ii) are satisfied, then $W_k(a)=1$.
The most important property of the witness function $W_k(a)$ is that
for a composite number $k\equiv k_C$ the number $t_{k_C}$ of 
witnesses $a$ s.t. 
$W_{k_C}(a)=0$ is 
\beq 
t_{k_C}\geq {3(k_C-1)\over 4},
\label{2}
\eeq
i.e., for a composite $k_C$ it is guaranteed that at least $3/4$ of
the $a<k_C$ are witnesses to $k_C$ \cite{rabin}- \cite{monier} (see
also ref. \cite{damgard}). 
For a prime number $k\equiv k_P$, instead, none of the $a$ is a
witness (i.e. $W_{k_P}(a)=1$ for all $1\leq a<k_P$).
Conversely, if, for an integer $1\leq a<k$ picked at random, 
one finds that $W_k(a)=1$,
then one can correctly declare $k$ to be prime with a probability 
$3/4$, while, if one finds $W_k(a)=0$, then $k$ can be declared composite with
certainty.

The classical randomized algorithm for the test of primality of $k$
heavily relies on this property of the witness function $W_k$
and proceeds as follows.
Given the number $k$ to be tested, 
one first picks up randomly $h$ numbers $a_i$ s.t.
$1\leq a_i<k$ ($i\in [1, h]$) and checks their witness function $W_k(a_i)$.
If $W_k(a_i)=0$ for {\it at least one} of the $a_i$, 
then $k$ is declared {\it composite}, while 
if $W_k(a_i)=1$ for {\it all} $a_i$, then $k$ is declared {\it prime}.
If $k$ is declared composite the test is always correct, but if $k$ is 
declared prime the test may fail with a probability 
(i.e., the probability of independently 
picking $h$ 'false' witnesses) smaller than $1/2^{2h}$ \cite{rabin}.
The computational complexity $S_{class}$ of the algorithm, defined as its 
running time as a function of the number of required operations, is 
polynomial in the number
of digits of $k$, i.e. $S_{class}\simeq O[h~ poly (\log k)]$.
\footnote[2]{One has to randomly generate (e.g., tossing $O[\log k]$ coins) 
$h$ numbers, taking $O[\log k]$ steps for each number,  
and to evaluate the witness function
for each of these numbers, taking another $O[~ poly(\log k)]$ steps
\cite{rabin}.}

We stress once more the main point leading to the good
performance of the classical randomized algorithm, i.e. the
{\it large gap between the number of witnesses in the cases when $k$ is
a prime and when it is a composite}.

\section{Quantum primality test}

Let us now present, at first, also as a useful 'warm-up' exercise,
a quantum algorithm to test the primality of a given number $k$
and compare it with the classical probabilistic one by Rabin.
The main idea underlying our quantum computation is the repeated
use of the counting algorithm COUNT originally introduced by Brassard et
al. \cite{brassard}.
The algorithm COUNT makes an essential use of two of the main
tools in quantum computation, i.e. Grover's unitary operation $G$
for extracting some elements from a flat superposition of
quantum states, and Shor's Fourier operation $F$ for extracting the
periodicity of a quantum state.
Grover's unitary transformation is given by $G=-WS_0WS_1$, where
the Walsh-Hadamard transform $W$ is defined as
\beq
W|a>\equiv {1\over \sqrt{k}}\sum_{b=0}^{k-1}(-1)^{a\cdot b}|b>
\label{w}
\eeq
(with $a\cdot b\equiv \sum_i a_ib_i ~\bmod 2$, $a_i(b_i)$ being the
binary digits of $a(b)$), 
$S_0\equiv I-2|0><0|$ and $S_1\equiv I-2\sum_{w}|w><w|$, which changes 
sign to the searched states $|w>$.
\footnote[3]{From here onwards, for simplicity, we use the compact notation 
according
to which, e.g. assuming $b\simeq O(k)$, we have $b= 2^jb_j+2^{j-1}
b_{j-1}+...+2^0b_0$ (with $j=[\log k]$) and $|b>$ itself is actually an
acronym for the tensor product of $j+1$ qubits, i.e. $|b>\equiv |b_j>
|b_{j-1}>\times ...\times |b_0>$.}
Shor's operation is, instead, given by the Fourier transform\footnote[4]{Note
that one can write the flat superposition as $W|0>=F|0>=\sum_a|a>/\sqrt{k}$.}
\beq
F|a>\equiv {1\over \sqrt{k}}\sum_{b=0}^{k-1}e^{2i\pi ab/k}|b>.
\label{f}
\eeq
Then, the COUNT algorithm can be summarized by the following
sequence of operations:

\vspace{1cm}
{\bf COUNT}:

~~1) $(W|0>)(W|0>)=\sum_m|m>\sum_a|a>$

~~2) $\rightarrow (F\otimes I)[\sum_m|m>G^m(\sum_a|a>)]$

~~3) $\rightarrow \mbox{measure} ~~|m>$
\vspace{1cm}

Since the amplitude of the set of the states $|w>$ after $m$ iterations of
$G$ on $|a>$ is a periodic function of $m$, the estimate of such a 
period by use of the Fourier analysis and the measurement of the
ancilla qubits in $|m>$ will give information on the size $t$ of this set,
on which the period itself depends.
The parameter $P$
determines both the precision of the estimate $t$ and the computational 
complexity of the COUNT algorithm (which requires $P$ iterations of
$G$).

Our quantum algorithm for the test of primality makes essential
use of the COUNT algorithm for estimating the number of
witnesses to the compositeness of $k$, and of $R\log P$ ancilla
qubits $|m_i>_i$ (with $m_i\in [0, P]$, $i\in [1, R]$ and $P$ is an
integer power of 2 to be determined later)
which are finally measured and which are necessary
in order to sharpen the constructive interference of the 'good' states.

We start with the tensor product of $R$ $|0>_i$ states with $\log P$ qubits, 
and one state $|0>$ with $\log k$ qubits, i.e.

\beq
|\psi_0>\equiv |0>_1....|0>_R|0>
\label{4}
\eeq
and then act on each of these states with a Walsh-Hadamard transform $W$
in order to obtain 

\beq
|\psi_1>\equiv {\sum_{m_1=0}^{P-1}|m_1>_1\over \sqrt{P}}.... 
{\sum_{m_R=0}^{P-1}|m_R>_R\over \sqrt{P}} {\sum_{a=0}^{k-1}|a>\over \sqrt{k}}. 
\label{5}
\eeq

We then act on the last $|a>$ state with the $|m_1>_1....|m_R>_R$-'controlled' 
Grover operation $G^m$ s.t.

\bea
|\psi_2>&\equiv &{\sum_{m_1=0}^{P-1}|m_1>_1\over \sqrt{P}}.... 
{\sum_{m_R=0}^{P-1}|m_R>_R\over \sqrt{P}}
\non \\
&\times & {\sum_{a=0}^{k-1}G^{m_1+....+m_R}
|a>\over \sqrt{k}}, 
\label{6}
\eea
where in $G$ we use $S_1\equiv I-2\sum_{W_k(a)=0}|a><a|$, which
changes sign to the witnesses of the compositeness of $k$.
\footnote[5]{A unitary 
transformation representing the witness function $W_k(a)$
can be easily obtained by defining the quantum $AND$ of the basic
operations in (i) and (ii) in eq. (\ref{1}), 
each of which can be evaluated in a time
which is polynomial in $\log k$.
For instance, one can first evaluate $h$ in $k-1\equiv 2^hl$ by reading the
highest qubit in $k-1$, and then build the state 
$|W^{(0)}_k(a)>....|W^{(h)}_k(a)>|W_k(a)>$, where $W^{(0)}_k(a)\equiv 
\Theta [a^{k-1}\bmod k]$, for $i\in [1, h]$ we have
$W^{(i)}_k(a)\equiv \Theta [\mbox{GCD}(a^{(k-1)/2^i}, k)-1 \bmod (k-1)+1]$,
with $\Theta[1]=1$ and $\Theta=0$ otherwise, and where
$|W_k(a)>\equiv |W^{(0)}_k(a)~\mbox{AND}....~W^{(h)}_k(a)>$.
The operator $S_1\sum_a|a>=-\sum_a(-1)^{W_k(a)}|a>$
can then be easily realized by tensoring the states $|a>$ with the ancilla 
qubit $|e>\equiv [|0>-|1>]/\sqrt{2}$ and acting with 
$U_{W_k}:|a>|e>\rightarrow |a>|e + W_k(a)\bmod 2>$.
All the operations leading to the evaluation of $W_k(a)$, except the last
for the phase change, have to be undone again, as usual, before acting with
$S_1$ and $G$.}

In the following we will assume that $P$ is at most $
\simeq O[poly (\log k)]$, so that
the steps required to compute the repeated Grover operations
\footnote[6]{Each of which has a computational complexity
$S_G\simeq O[poly(\log k) +\log k]\simeq O[poly(\log k)]$, 
the first term being for the 
quantum parallel evaluation of the witness functions, and the second term 
for the evaluations of the $W$ and $S_0$ transforms.} 
$G^{m_1+....+m_R}$ is polynomial in $\log k$.

We define the quantities

\beq
\sin\theta_k\equiv \sqrt{t_k\over k}
\label{7}
\eeq
and

\bea
k_{m_1....m_R}&\equiv & \sin [2(m_1+....+m_R)+1]\theta_k
\non \\
l_{m_1....m_R}&\equiv & \cos [2(m_1+....+m_R)+1]\theta_k,
\label{7a}
\eea
where $t_k$ is the number of $a$ s.t. $1\leq a<k$ and $W_k(a)=0$, 
and the states

\bea
|B_1>_k&\equiv & {1\over \sqrt{t_k}}\sum_{W_k(a)=0}|a>
\non\\
|B_2>_k&\equiv & {1\over \sqrt{k-t_k}}\sum_{W_k(a)=1}|a>,
\label{8}
\eea
such that we can simplify eq. (\ref{6}) by use of

\bea
{1\over \sqrt{k}}\sum_{a=0}^{k-1}G^{m_1+....+m_R}|a>&=&k_{m_1....m_R}|B_1>_k
\non \\
&+&l_{m_1....m_R}|B_2>_k.
\label{9}
\eea

Next we apply Shor's Fourier transform on each of the $R$ ancilla states
$|m_i>_i$ 
in order to extract the periodicity $\theta_k$ (and, therefore, via eq.
(\ref{7}), the number of witnesses $t_k$) which is hidden in the 
amplitudes $k_{m_1....m_R}$ and $l_{m_1....m_R}$, i.e. we transform $|\psi_2>$ into

\bea
|\psi_3>&\equiv &{\sum_{m_1, l_1=0}^{P-1}e^{2i\pi l_1m_1/P}|l_1>_1\over P}.... 
\non \\
&\times &{\sum_{m_R, l_R=0}^{P-1}e^{2i\pi l_Rm_R/P}|l_R>_R\over P}
\non \\ 
&\times &[k_{m_1....m_R}|B_1>_k+
l_{m_1....m_R}|B_2>_k].
\label{10}
\eea

After some elementary algebra, eq. (\ref{10}) can be rewritten as 

\bea
|\psi_3>&\equiv &{1\over 2}\sum_{l_1,...l_R=0}^{P-1}|l_1>_1....|l_R>_R
e^{-i\pi (l_1+....+l_R)P}
\non \\
&\times & \biggl [e^{i\pi f_k^{(R)}}\prod_{i=1}^{R}
s_{l_i+}^{(P)}(-i|B_1>_k+|B_2>_k)
\non \\
&+&e^{-i\pi f_k^{(R)}}\prod_{i=1}^{R}
s_{l_i-}^{(P)}(i|B_1>_k+|B_2>_k)\biggr ],
\label{11}
\eea
where we have introduced the following quantities,

\bea
f_k&\equiv & {P\theta_k\over \pi}~~~~ ;~~~~0\leq f_k\leq {P\over 2}
\non \\
f_k^{(R)}&\equiv & f_k\left [R+{(1-R)\over P}\right ]
\label{12}
\eea
and

\beq
s_{l_i\pm}^{(P)}\equiv {\sin\pi (l_i\pm f_k)\over 
P\sin{\pi(l_i\pm f_k)\over P}}.
\label{13}
\eeq

In particular, when counting the witnesses for a given $k$, we have
two different possibilities: either $k$ is a prime, in which case we have 
that $t_{k_P}=0$ and therefore $\theta_{k_P}=f_{k_P}=0$; or $k$ is a composite,
for which we have that $t_{k_C}\geq 3{k_C}/4$ and $\theta_{k_C}\geq \pi/3$,
implying that $P/3\leq f_{k_C}\leq P/2$.

Going back to eq. (\ref{11}), we can see that, in the case when $k$ is 
a prime, $G$ effectively acts as an identity operator, so that
$|\psi_3>$ simplifies to 

\beq
|\psi_3>\rightarrow |0>_1....|0>_R|B_2>_k~~~~;~~~~ \mbox{when $k=k_P$}.
\label{14}
\eeq


On the other hand, when $k$ is a composite, almost all of the ancilla qubits
in $|\psi_3>$ will be in a state different from $|0>_1....|0>_R$.
In fact, the probability of finally measuring $|0>_1....|0>_R$ when
$k$ is composite is

\bea 
P(|0>_1....|0>_R)\biggr |_{k_C}&=& (\alpha_k)^{2R}\biggr |_{k_C}\equiv 
\left ({\sin \pi f_k\over P\sin {\pi f_k\over P}}\right )^{2R}\biggr |_{k_C}
\non \\
&\leq &\left ({2\over \sqrt{3} P}\right )^{2R}
\simeq O[P^{-2R}],
\label{15}
\eea
since we have $f_{k_C}\geq P/3$.

Summarizing the above results, our quantum algorithm for testing
the primality of a given number $k$ is probabilistic in the following sense:
if in the final measurement process of the $R\log P$ ancilla qubits
we obtain a state with {\it at least one} of the qubits different
from $|0>$, we can declare with {\it certainty} that the number $k$ is
a {\it composite}; on the other hand, if {\it all} the ancilla qubits are
in the state $|0>$, we can claim with an {\it error probability smaller
than} $O[P^{-2R}]$ that the number $k$ is a {\it prime}. 

The use of $R\log P$ ancilla qubits and the repeated application of the
Fourier transforms is made
in order to sharpen the constructive interference effects at the 
basis of the measurement of the period $\theta_k$ and, in this sense, 
it can be seen as the quantum analogue of the multiple random tests
used in the classical primality algorithm by Rabin.
Moreover, our algorithm, provided that the final measurement step is
omitted, is clearly unitary and fully reversible, and as such it can
be used as an intermediate unitary transform inside a larger and
more complicated algorithm.
The strength of this quantum algorithm then critically relies on
the use of the superposition and entanglement of states, and above
all on the existence of a {\it gap} between the cardinalities
of different sets of the domain of a given test function (in the case of the 
function $W_k(a)$ the domain is divided in the set of states
with $W_k=1$, $f_k=0$, and those with $W_k=0$, $f_k\geq P/3$).
\footnote[7]{Taking 
$f_{good}=0$ (i.e., $\alpha_{good}=1$) and, more in general,
$f_{bad}=\xi P/2$ with $\xi\in (0, 1]$, and requiring a success probability 
exponentially close to one, we have to choose, e.g., $\alpha_{bad}\leq  
(P\sin \pi\xi/2)^{-1}<1/2$ in eq. (\ref{15}).
Thus, actually even small values of $\xi$ (but $\xi > [poly(\log N)]^{-1}$)
are good enough for the quantum test to be 
sufficiently reliable, provided one takes $P\geq O[\xi^{-1}]$.} 
We will show other and more interesting problems where these properties
can be fruitfully exploited in section 4 and in the appendix.  

The computational complexity of the quantum algorithm can be written as 
$S_{quant} \simeq O[\log k+R\{\log P +(\log P)^2+PS_G\}]\simeq O[RPS_G]$,
\footnote[8]{The first term is for the construction of the flat superposition
$\sum_a|a>$, while the other terms arise, in the order from left to right, 
from the evaluations of the $R$ flat superpositions of the
ancilla states $\sum_{m_i}|m_i>_i$ ($i\in [1, R]$),
the $R$ Fourier transforms on the same states
and the operation $G^{\sum m_i}$, requiring $PR$ repetitions of the basic 
block $G$.} with the number of steps required for $G$  given by 
$S_G\simeq O[poly(\log k)]$ (see footnote below eq. \rf{6}),
so that we obtain $S_{quant}\simeq O[R ~poly(\log k)]$.
\footnote[9]{One might observe that, in fact, it is not necessary to
use COUNT for primality testing, but simply build the state
$\left [ \sum_{a}|a>|W_k(a)>\right ]$ (for the bases $a$ of the integer $k$),
measure the ancilla qubit $|W_k(a)>$ and repeat the procedure $h$ times.
This would imply the same error probability $P_e\simeq 2^{-2h}$ (after
$h$ trials), computational complexity $S_{quantum}\simeq O[poly (\log k )]$ and
memory space required $M=\log k +1$ as in our algorithm with the choice
$R=1, P\simeq O(1)$ repeated $h$ times, and in the classical algorithm
of ref. \cite{rabin}.
However, this method of repeated trials would not make the subroutine 
for primality testing unitary and reversible.
To achieve this goal, one might instead consider (see, e.g., ref. 
\cite{bennett}) the initial state
$\left [ \sum_{a_1}|a_1>|W_k(a_1)>\right ]....\left [ \sum_{a_R}|a_R>
|W_k(a_R)>\right ]$ and parallely check for the $R$ ancilla qubits 
$|W_k(a_i)>$, for which the error probability would be 
$P_{e, bennett}\simeq 2^{-2R}$, the computational complexity would be
$S_{quantum, bennett}\simeq O[ R (poly (\log k ))]$ and the memory
space used would be $M_{bennett}=R(\log k +1)$.
With the choice $P\simeq O(1)$ and $R\simeq O(\log k)$, however, our
algorithm has the same error probability and computational complexity,
but requires logarithmically less memory space.
This makes explicit the advantage of exploiting the interference among
quantum states which is inherent in our method via the use of COUNT.}

\section{Counting $k_P<N$ and the prime number theorem}

One of the problems in which the quantum algorithm of the previous section
can be explicitly used, as the basic block of another more complex
unitary operation, is the case of the testing of the so called
'prime number theorem' (see, e.g., ref. \cite{ribenboim} and references
therein), according to which
the total number $t_N$ of primes $k_P$ smaller than a given number $N$
is given by the formula 
\beq
t_N\equiv \pi(N)\simeq {N\over \log N}
\label{t}
\eeq

Our quantum algorithm essentially consists of a sub-loop which checks for the
primality of a given $k<N$ by counting its witnesses, a main loop for the
counting of primes less than $N$, and a final measurement of some ancilla 
qubits.
More in details, we can schematically summarize the main operations 
in the following steps:

\vspace{1cm}
{\bf MAIN-LOOP}:

~~{\it Count $\sharp \{k | k=k_P< N \}$ using {\bf COUNT} with
$G\rightarrow {\tilde G}$ and $S_1\rightarrow {\tilde S}_1\equiv 
1-2\sum_{k_P}|k_P><k_P|$ (parameter $Q$)} 

{\bf SUB-LOOP}:

~~{\it Parallel primality tests $\forall ~k<N$ (parameter $P$) and
(approximate) construction of ${\tilde S}_1$}
\vspace{1cm}

Let us start from the SUB-LOOP of the algorithm first.
The unitary transform ${\tilde S}_1$ to be computed
should approximate
with a high level of accuracy the following basic operation 

\beq
{\tilde S}_1~:~{1\over \sqrt{N}}\sum_{k=0}^{N-1}|k>\rightarrow 
{1\over \sqrt{N}}\sum_{k=0}^{N-1}(-1)^{F_k}|k>,
\label{22}
\eeq
where $F_k\equiv 1$ for a prime $k=k_P$ and $F_k\equiv 0$ for a
composite $k=k_C$.

In order to construct such an ${\tilde S}_1$, 
we start from the flat superposition
of states $|k>$ tensored with two ancilla states $|0>_P$ and $|0>_k$ with, 
respectively, $\log P$ and $\log k$ qubits, i.e.

\beq
|{\bar \psi}_0>\equiv {1\over \sqrt{N}}\sum_{k=0}^{N-1}|k>|0>_P|0>_k,
\label{23}
\eeq
and we act on the first two states ($|k>$ and $|0>_P$)
in eq. (\ref{23})
with an $F$ transform and a $|k>$-'controlled' $F$ operation, respectively, 
to get

\beq
|{\bar \psi}_1>\equiv {1\over \sqrt{N}}\sum_{k=0}^{N-1}|k>
{\sum_{m=0}^{P-1}|m>_P\over
\sqrt{P}}{\sum_{a=0}^{k-1}|a>_k\over \sqrt{k}}
\label{24}
\eeq
(with $P\simeq O[poly(\log N)]$).
Then, as usual, we operate with a $|m>_P$-'controlled' Grover 
transform $G^m$ on the last ancilla states $|a>_k$ followed by a Fourier 
transform $F$ on $|m>_P$, obtaining

\bea
|{\bar \psi}_2>&\equiv &{1\over \sqrt{N}}\sum_{k=0}^{N-1}|k>\biggl 
[\alpha_k|0>_P|A_0>_k
\non \\
&+&\sum_{m=1}^{P-1}\phi_{m}|m>_P|A_m>_k\biggr ],
\label{25}
\eea
where 

\bea
|A_0>_k&\equiv &\sin \pi f_k|B_1>_k+\cos \pi f_k|B_2>_k
\non \\
|A_m>_k&\equiv &[e^{-i\pi f_k}s_{m -}^{(P)}(i|B_1>_k+|B_2>_k)
\non \\
&+&e^{i\pi f_k}s_{m +}^{(P)}(-i|B_1>_k+|B_2>_k)]/2
\label{26}
\eea
while $\sin\theta_k$, $|B_{1,2}>_k$, $f_k$, $s_{m\pm}^{(P)}$ and 
$\alpha_k$ have been 
defined, respectively, in eqs. (\ref{7}), (\ref{8}), (\ref{12}), (\ref{13}) 
and (\ref{15}), and the phase $\phi_{m} \equiv \exp[i\pi m(1-1/P)]$.

We now act with the phase change operator $S_0$
on the first ancilla state $|\cdot>_P$, and then undo again all the 
previous operations
($F$, $G^m$, and the two initial $F$s) finally obtaining  the state

\beq
|{\bar\psi}_3>\equiv {1\over \sqrt{N}}\sum_{k=0}^{N-1}|k>[|0>_p|0>_k-2
|C_1>_{P, k}],
\label{27}
\eeq
with 

\bea
|C_1>_{P, k}&\equiv &\alpha_k \sum_{n, r=0}^{P-1}\sum_{b=0}^{k-1}
e^{-2i\pi nr/P}[(\mbox{Im}~Z_{k, r})B_{1, k}(b)
\non \\
&+&(\mbox{Re}~Z_{k, r})B_{2, k}(b)]{|n>_P|b>_k\over P\sqrt{k}}
\label{28}
\eea
and

\bea
B_{1, k}(b)&\equiv & {\sum_{W_k(a)=0}e^{-2i\pi ab/k}
\over\sqrt{t_k}}
\non \\
B_{2, k}(b)&\equiv & {\sum_{W_k(a)=1}e^{-2i\pi ab/k}
\over \sqrt{k-t_k}}
\non \\
Z_{k, r}&\equiv & e^{i\pi f_k}e^{-2i\pi f_k r/P}.
\label{29}
\eea

Noting the properties that ${~}_{P, k}<C_1|C_1>_{P, k}=\alpha_k^2$ and, 
for a prime
$k_P$, $\alpha_{k_P}=1$, with $|C_1>_{P, k_P}=|0>_P|0>_{k_P}$, we can also
rewrite eq. (\ref{27}) as

\beq
{\tilde S}_1|{\bar \psi}_0>=|{\bar\psi}_3>\equiv |\Psi>+|E>,
\label{30}
\eeq
where

\bea
|\Psi>&\equiv & {1\over \sqrt{N}}
\sum_{k=0}^{N-1}(-1)^{F_k}|k>|0>_P|0>_k
\non \\
|E>&\equiv & -{2\over \sqrt{N}}\sum_{k=k_C}|k>|C_1>_{P, k},
\label{31}
\eea
which realize, as wanted, the operation ${\tilde S}_1$ of eq. (\ref{22}),
with the norm of the correction term $|E>$ upper bounded by 

\bea
<E|E>&=&{4\over N}\sum_{k=k_C}\alpha_k^2
\non \\
&\leq &4\left ({2\over \sqrt{3}P}
\right )^2\simeq O[P^{-2}].
\label{32}
\eea

Defining, in a symbolic notation, the sequence of operations

\beq
U_1\equiv F~[CTRL_{|m>_P}(G)]~[CTRL_{|k>}(F)]~F,
\label{33}
\eeq
we have, in fact,
\beq
{\tilde S}_1\equiv U^{\dagger}_1S_0U_1.
\label{34}
\eeq

Let us now consider the MAIN-LOOP of the algorithm, i.e. that counting 
the total number of $k_P<N$.
Grover's transform $\tilde G$ entering this part of the algorithm can then
be written as

\beq
\tilde G\equiv U_2~{\tilde S}_1~~~~;~~~~U_2\equiv -W^{(k)}S_0^{(k)}W^{(k)},
\label{35}
\eeq
and with the caveat that now the operations $W^{(k)}$ and $S_0^{(k)}$ appearing
in the operator $U_2$ of eq. (\ref{35}) are acting on the states $|k>$, 
and that the states to be counted finally are those with $k=k_P$.

Defining as usual 

\beq
\sin\theta_N\equiv \sqrt{t_N\over N}
\label{37}
\eeq
and the 'good' and 'bad' states, respectively, as

\bea
|G>&\equiv & {\sum_{k_P}|k>|0>_P|0>_k\over \sqrt{t_N}}
\non \\
|B>&\equiv & {\sum_{k_C}|k>|0>_P|0>_k\over \sqrt{N-t_N}},
\label{38}
\eea
we have then

\bea
{\tilde S}_1|G>&=&-|G>
\non \\
{\tilde S}_1|B>&=&|B>+\sec\theta_N |E>. 
\label{40}
\eea

Consequently, we can derive a formula for the iteration of
the operator $\tilde G$ acting on the state $|{\bar \psi}_0>$, i.e.

\beq
{\tilde G}^n|{\bar \psi}_0>=G^n|{\bar \psi}_0>+|E_n>,
\label{41}
\eeq
with

\beq
|E_n>\equiv\sec\theta_N\left [\sum_{j=1}^n ~l_{n-j}{\tilde G}^{j-1}\right ]
U_2|E>,
\label{42}
\eeq
where we have again used the variables 
$k_m\equiv \sin(2m+1)\theta_N$ and $l_m\equiv \cos(2m+1)\theta_N$, the
formulas
\bea
|{\bar \psi}_0>&=&\sin \theta_N|G>+\cos \theta_N|B>
\non \\
|\Psi>&=&-\sin \theta_N|G>+\cos \theta_N|B>
\label{39}
\eea
and eq. (\ref{40}).

We have now all the building blocks necessary to proceed
with the construction of the quantum algorithm counting the
number of $k_P$ s.t. $k_P<N$.
We start from $|{\bar \psi}_0>$ given by formula (\ref{23}) and tensor it
with a flat superposition of ancilla states $|m>_Q$ with $\log Q$ qubits, i.e.

\beq
|{\bar \psi}_4>\equiv {1\over \sqrt{Q}}\sum_{m=0}^{Q-1}|m>_Q|{\bar \psi}_0>
\label{43}
\eeq
(with $Q$ an integer power of 2 to be chosen later of $O[poly (\log N)]$), 
then we act on $|{\bar \psi}_0>$ 
with the $|m>_Q$-'controlled' ${\tilde G}^m$
and with $F$ on $|m>_Q$, getting

\bea
|{\bar \psi}_5>&\equiv &{1\over Q}\sum_{m, n=0}^{Q-1}e^{2i\pi mn/Q}|n>_Q[
k_m|G>+l_m|B>
\non \\
&+&|E_m>]={1\over 2}\sum_{n=0}^{Q-1}e^{i\pi n(1-1/Q)}|n>_Q
[e^{-i\pi f_Q}s^{(Q)}_{n-}
\non \\
&\times &(i|G>+|B>)+e^{i\pi f_Q}s^{(Q)}_{n+}(-i|G>+|B>)]
\non \\
&+&{1\over Q}\sum_{m, n=0}^{Q-1}e^{2i\pi mn/Q}|n>_Q|E_m>,
\label{44}
\eea
where 

\beq
f_Q\equiv Q\theta_N/\pi
\label{45b}
\eeq
and $s^{(Q)}_{n\pm}$ are defined in eq. (\ref{13}).

Now, the last step of the algorithm consists in measuring the
value of the state $|\cdot >_Q$ in $|{\bar \psi}_5>$.
Using the expected estimate that $\theta_N\simeq O[1/\sqrt{\log N}]$, which
gives $f_Q\simeq O[Q/\sqrt{\log N}]$, and by choosing 

\beq
Q\simeq O[(\log N)^{\beta}]~~~~;~~~~\beta>1/2, 
\label{q}
\eeq
we get the ansatz $1< f_Q<Q/2-1$ of ref. \cite{brassard}.
%
%
Then it can be easily shown, exactly as in ref. \cite{brassard}, that the 
probability $W$ to
obtain any of the states $|f_->_Q$,  $|f_+>_Q$,  $|Q-f_->_Q$ or  $|Q-f_+>_Q$  
(where $f_-\equiv [f_Q]+\delta f$ and $f_+\equiv f_- +1$, with $0<\delta f<1$)
in the final measurement is given by

\beq
W\geq {8\over \pi^2}-|W_{E_n}|,
\label{46}
\eeq
where $W_{E_n}$ is the contribution coming from terms involving $|E_n>$,
and whose explicit form we omit here for simplicity.

Using the upper bound $<E_n|E_n>\leq O[n^2]<E|E>$\hfill\break
and choosing

\beq
P\simeq O[(\log N)^{\gamma}]~~~~;~~~~\gamma>\beta ,
\label{p}
\eeq
from eq. (\ref{32}) we then get the estimate\footnote[10]{The condition 
(\ref{p})
is sufficient but not necessary in order to have $|W_{E_n}|\ll 1$.
In fact, one can also choose $P=cQ$, provided that the constant $c\ll 1$.} 

\beq
<E_n|E_n>~\leq O[(\log N)^{-2(\gamma-\beta)}]\ll 1
\label{47}
\eeq
which, substituted in the formula for $W_{E_n}$ and, then, 
in eq. (\ref{46}), finally gives the estimate

\beq
W\geq {8\over \pi^2} \{ 1-O[(\log N)^{-(\gamma-\beta)}]\}.
\label{48}
\eeq

This means that with a high probability we will always be able to
find one of the states $|f_{\pm}>_Q$ or $|P-f_{\pm}>_Q$ and, therefore,
to evaluate the number $t_N$ from eqs. (\ref{37}) and (\ref{45b}).
\footnote[11]{We note
that, as an alternative of choosing $P$ as in eq. (\ref{p}), one could 
also repeat the counting
algorithm a sufficient number of times, as we did in the previous section
(see eq. (\ref{5})), in order to reduce the 'error' probability $W_{E_n}$.}

Of course, as explained in ref. \cite{brassard}, since in general $f_Q$
is not an integer, the measured ${\tilde f}_Q$ 
will not match exactly the true value of $f_Q$, and we will have some
errors.
In particular, defining ${\tilde t}_N\equiv N\sin^2{\tilde \theta}_N$,
with ${\tilde \theta}_N={\tilde \theta}_N({\tilde
f}_Q)$, we have for the error over $t_N$ 
the estimate\cite{brassard}

\bea
|\Delta t_N|_{exp}&\equiv &|{\tilde t}_N-t_N|\leq\pi{N\over Q}
\left [{\pi \over Q}+2\sqrt{t_N\over N}\right ]
\non \\
&\simeq &O[N(\log N)^{-\beta-1/2}].
\label{49}
\eea

On the other hand, if we want to check the theoretical formula 
$t_N\equiv \pi(N)$ up to some power $\delta>0$ in $\log N$, i.e. with

\beq
|\Delta t_N|_{th}\simeq O[N(\log N)^{-\delta -1}],
\label{50}
\eeq
we have to impose that the measuring error over $t_N$ is smaller
than the precision required for testing $t_N$, i.e. we should have
$|\Delta t_N|_{exp}< |\Delta t_N|_{th}$, which
can be satisfied provided that

\beq
\beta >\delta +1/2.
\label{51}
\eeq

The computational complexity of the quantum algorithm can be written as
$S_Q\simeq O[\log N +\log Q +(\log Q)^2 +Q(\log N + S_1)]$,
where for the SUB-LOOP we have $S_1\simeq O[\log N+\log P + PS_G + 
(\log P)^2]$, and which, using eqs. (\ref{q}), (\ref{p}) and
$S_G\simeq O[poly(\log N)]$, finally gives the {\it polynomial} complexity 
$S_Q\simeq O[QPS_G]\simeq O[poly (\log N)]$.

As noted in ref. \cite{brassard}, moreover, one can 
further minimize the errors by successive repetitions of the whole algorithm.
In particular, it is easily seen that the success probability $W$ can
be boosted exponentially close to one and an exponential accuracy
can be achieved by repeating the whole algorithm many times and using
the majority rule, still leaving the whole algorithm for the test of the 
'prime number theorem' polynomial in $\log N$.\footnote[12]{The average 
computational complexity can be further, slightly reduced by use of parallelism
and anticipate measurements [6, 12].}

We conclude by stressing, once again, that the power of our quantum
probabilistic methods essentially relies on:
the {\it gap} between the cardinalities of the domains of the test function
$W_k$;
the fact that the probability to obtain any of the states $f^{\pm}$ or
$Q-f^{\pm}$ is bigger than 1/2 (which is true provided that the 'error'
terms $|E_n>$ have sufficiently small amplitude); that, finally, the
error over the estimate of $t_N$ is smaller than the precision we
need (the last two conditions being strongly dependent on the choice of
$Q, Q/P, \sqrt{k/t_k}$ and $\sqrt{N/t_N}$).

\section{Discussion}

In this paper we have shown a method to build a quantum version of 
the classical probabilistic algorithms $\grave{a}$ la Rabin.
Our quantum algorithms make essential use of some of the basic
blocks of quantum networks known so far, i.e. Grover's
operator for the quantum search of a database, Shor's Fourier
transform for extracting the periodicity of a function and their
combination in the counting algorithm of ref. \cite{brassard}.
The most important feature of our quantum probabilistic algorithms
is that the coin tossing used in the correspondent classical probabilistic 
ones is replaced here by a unitary and reversible
operation, so that the quantum algorithm can even be used as a subroutine
in larger and more complicated networks. 
In particular, we described polynomial time algorithms for studying 
some problems in number theory, e.g. a primality test, the 
'prime number theorem' and a conjecture concerning a certain 
distribution of couples of primes.
Our quantum algorithm may also be useful for other similar tests and counting
problems if there exists a classical probabilistic algorithm which somehow 
can guarantee a good success probability (e.g.,
problems related to the distribution of 
primes and pseudoprimes in number theory etc..).\setcounter{footnote}{1}
\footnote[1]{Other quantum algorithms dealing 
with problems in number theory,
such as integer factoring, finding discrete logarithms or a Pocklington-Lehmer
primality test can be found, respectively, in refs. \cite{shor} and
\cite{chau}.
The extent to which the algorithm presented in the latter work, however, 
can be actually used as an efficient primality test is very questionable
for us.}
It is well known that in a classical computation one can count, by using
Monte-Carlo methods, the cardinality of a set which satisfies some conditions,
provided that the distribution of the elements of such a set is assumed to 
be known (e.g., homogeneous).
One further crucial strength and novelty of our algorithm is also in the 
ability of efficiently and successfully solve problems where such a knowledge 
or regularities may not be present.

\appendix
\section{Testing Hardy and Littlewood's conjecture}

A very similar procedure can be followed for testing Hardy and Littlewood's 
conjecture \cite{hardy} that the number, which we call $r_2(2N)$, of 
the possible representations of an even number $2N$ as the sum of two primes
$k_P, l_P<2N$ such that $k_P+l_P=2N$, should be given by the 
asymptotic formula (modulo $O[\log\log N]$ factor corrections \cite{hardy})

\beq
r_2(2N)\biggl |_{th}
\simeq O\left [{N\over (\log N)^{\mu}}\right ]~~~~;~~~~\mu \simeq 2.
\label{a1}
\eeq

The quantum algorithm for counting such couples and testing
the conjecture  can be built starting from the state

\bea
|\phi_0>&\equiv &{1\over \sqrt{2N}}\sum_{k=0}^{2N-1}|k>|2N-k>
\non \\
&\times &|0>_1|0>_2|0>_3|0>_4
\label{a2}
\eea
where the four ancilla states $|0>_i$ have, respectively, 
$\log P$ qubits for $i=1,2$ and $\log (2N-1)$ qubits for $i=3,4$, 
and then, as done in section 4, constructing an operator
$S^{\prime}_1$ such that

\beq 
S^{\prime}_1:~|\phi_0>\rightarrow |\Phi_0>+|E^{\prime}>,
\label{a3}
\eeq
with the main contribution $|\Phi_0>$ given by the wanted phase change for the
'good' states $|k_P>|(2N-k)_P>$, i.e.

\bea
|\Phi_0>&\equiv &{1\over \sqrt{2N}}\sum_{k=0}^{2N-1}(-1)^{G_k}|k>|2N-k>
\non \\
&\times &|0>_1|0>_2|0>_3|0>_4,
\label{a4}
\eea
where $G_k\equiv 1$ for the 'good' couples $|k_P>|(2N-k)_P>$ and 
$G_k\equiv 0$ for all the other couples, and $|E^{\prime}>$ is a
correction whose amplitude should be negligible with respect to
that of $|\Phi_0>$.
This result is achieved, once again, starting from $|\phi_0>$ and
acting with an $F$ on the first two ancilla states, 
with a $|k>$- and $|2N-k>$-'controlled' operator $F$, respectively, on 
each of the last two ancilla states, then operating
with a controlled-$G$ transform on each of the last two ancilla states 
and with an $F$ transform on each of the first two ancilla states, 
inverting the phase of the state $|0>_1|0>_2|0>_3|0>_4$ and 
finally undoing the previous operations again.
Doing so, one obtains as promised eqs. (\ref{a3}-\ref{a4}), 
with the explicit formula
for the state $|E^{\prime}>$ given by (we omit all the algebraic details of
the derivation for the sake of simplicity)

\bea
|E^{\prime}>&\equiv &-{2\over \sqrt{2N}}\sum_{k^{\prime}=0}^{2N-1}
\alpha_{k^{\prime}}\alpha_{(2N-k)^{\prime}}|k^{\prime}>|(2N-k)^{\prime}>
\non \\
&\times &|C_2>_{k^{\prime}},
\label{a5}
\eea
where the sum $k^{\prime}$ is over all couples except $|k_P>|(2N-k)_P>$,
and the norm of the state $|C_2>_{k^{\prime}}$ (a certain tensor product of
the four ancilla states) is of $O(1)$.

To count the number of 'good' couples one has to repeat exactly the
same steps as described from eq. (\ref{35}) (with ${\tilde S}_1\rightarrow
S_1^{\prime}$) to eq. (\ref{49}) (with $t_N\rightarrow r_2(2N)$) 
in section 4,
and the final result is that, again, the expected theoretical
behaviour of $r_2(2N)$ can be tested up to exponential accurarcy in
a polynomial number of steps.
In particular, if one takes 

\bea
Q&\simeq &O[(\log N)^{\rho}]~~~~;~~~~\rho >\mu/2
\non \\
P&\simeq &O[(\log N)^{\sigma}]~~~~;~~~~\sigma>\rho,
\label{a6}
\eea
so that $1<f_Q<Q/2-1$ in the MAIN-LOOP of the 
algorithm, that $<E_n^{\prime}|E_n^{\prime}>\leq 
O[P^{-2(\sigma -\rho)}]$ and $W\geq 8/\pi^2\{1-O[P^{-(\sigma -\rho)}]\}$, if 
we want to test the $2N$-dependence of $r_2(2N) $ with a theoretical error 

\beq
|\Delta r_2(2N)|_{th}\simeq O[N(\log N)^{-\mu -\nu}]~~~~;~~~~\nu >0,
\label{a7}
\eeq
the condition $|\Delta r_2(2N)|_{exp}< |\Delta r_2(2N)|_{th}$ requires that

\beq
\rho>\mu/2+\nu .
\label{a8}
\eeq

The evaluation of the quantum computational complexity of the
algorithm can be shown to be polynomial in the number $2N$ 
in a fashion similar to section 4.

Finally, we should comment that the method used could be easily 
extended to other similar counting problems, such as the case, e.g., 
when one wants 
to check that a given integer is the sum of more than two
primes or their certain powers etc...
In fact, provided that one is able to check with this quantum
algorithm that {\it any} odd integer less than $N_0=10^{43000}$ (!) \cite{chen}
can be written as the sum of three primes, then one might also have a
numerical (although probabilistic) tool to prove with a {\it polynomial}
computational complexity the weaker version
of a famous Goldbach's conjecture, i.e. that every odd integer $N>5$ is 
the sum of three primes (see, e.g., ref. \cite{ribenboim}).
 
\vspace{33pt}
\noindent {\Large \bf Acknowledgements}

\bigskip
A.H.'s research was partially supported by the Ministry of Education, 
Science, Sports and Culture of Japan, under grant n. 09640341.
A.C.'s research was supported by the EU under the Science and Technology
Fellowship Programme in Japan, grant n. ERBIC17CT970007; he also thanks
the cosmology group at Tokyo Institute of Technology for the kind hospitality 
during this work.
Both authors would like to thank Prof. T. Nishino for introducing them
to Rabin's primality test, Prof. N. Kurokawa for helpful discussions and
the referees for pointing out some useful comments.

\end{document}